\documentclass{aa}
\usepackage{aas_macros}
\usepackage{graphicx, epsfig}
\usepackage{amssymb}
\usepackage{natbib}
\bibpunct{(}{)}{;}{a}{}{,}

\newcommand{\lam}{$\lambda$}

\def\element#1#2#3{\ensuremath{\mathrm{^{#1}#2}\,\textsc{#3}}} 
\def\ion#1#2{\ensuremath{\mathrm{#1^{#2}}}} 
 
\newcommand{\oiii}{[\element{}{O}{iii}]}
\newcommand{\oii}{[\element{}{O}{ii}]}
\newcommand{\oi}{[\element{}{O}{i}]}
\newcommand{\sii}{[\element{}{S}{ii}]}
\newcommand{\siii}{[\element{}{S}{iii}]}
\newcommand{\hei}{\element{}{He}{i}}
\newcommand{\heii}{\element{}{He}{ii}}

\newcommand{\hel}{\element{4}{He}}

\newcommand{\hii}{\element{}{H}{ii}}

\newcommand{\clo}{{\sc Cloudy}}
\def\teff{\ensuremath{{T_{\mbox{\tiny eff}}}}}

\def\icf{\ensuremath{\mathit{icf}}}
\def\icfx{\ensuremath{\mathit{icf^{*}}}}
\def\tcf{\ensuremath{\mathit{tcf}}}

\DeclareRobustCommand{\ionsect}[2]{%
\relax\ifmmode
\ifx\testbx\f@series
{\ensuremath{#1\,\mathsc{#2}}}\else
 {\ensuremath{\mathrm{#1\,\mathsc{#2}}}}\fi
 \else\textup{#1\,{\mdseries\textsc{#2}}}%
 \fi}

\begin{document}

   \title{Systematic uncertainties in the determination\\ of the primordial
   $^{4}$He abundance}
   \author{D. Sauer%
          \and
          K. Jedamzik%
          } 
        \institute{Max-Planck-Institut f\"ur Astrophysik,
          Karl-Schwarzschild-Str. 1, 85741 Garching, Germany}

        \offprints{\\ D. Sauer, \email{dsauer@mpa-garching.mpg.de} or\\
        K. Jedamzik, \email{jedamzik@mpa-garching.mpg.de}} 
        \date{Received  / Accepted }

        \abstract{The primordial helium abundance $Y_p$ is commonly
          inferred from abundance determinations in low-metallicity
          extragalactic \hii-regions. Such determinations may be
          subject to systematic uncertainties that are investigated
          here. Particular attention is paid to two effects:
          \icf{}-corrections for ``imperfect'' ionization structure
          leading to significant amounts of (unobservable) neutral
          helium or hydrogen and ``\tcf{}''-corrections due to
          non-uniform temperature. Model \hii-regions with a large
          number of parameters are constructed and it is shown that
          required corrections are almost exclusively functions of two
          physical parameters: the number of helium- to hydrogen-
          ionizing photons in the illuminating continuum
          $Q(\mbox{\ion{He}{0}})/Q(\mbox{\ion{H}{0}})$, and the ratio
          of width to radius $\delta r_{S}/r_{S}$ of the Str\"omgren
          sphere.  For clouds of sufficient helium-ionizing photons
          $Q(\mbox{\ion{He}{0}})/Q(\mbox{\ion{H}{0}}) \gtrsim 0.15$
          and non-negligible width of the Str\"omgren sphere a
          significant overestimate of helium abundances may result.
          Such clouds show radiation softness parameters in the range
          $-0.4 \lesssim \log\,\eta \lesssim 0.3$ coincident with the
          range of $\eta$ in observed \hii-regions. Existing data of
          \hii-regions indeed seem to display a correlation which is
          consistent with a typical $\sim 2-4\%$ overestimate of
          helium abundances due to these effects.  In case such an
          interpretation prevails, and in the absence of other
          compensating effects, a significant downward revision of
          $Y_p$ may result. It is argued that caution should be
          exercised regarding the validity of commonly quoted error
          bars on $Y_p$.  
           } 
        \authorrunning{Sauer \& Jedamzik}
        \titlerunning{Systematic Uncertainties in the Determination of
          $Y_{P}$} \maketitle



\section{Introduction}
\label{sec:intro}

The observational determination of the primordial light element
abundances provides an important key test for the validity of the
standard model of Big Bang nucleosynthesis and the cosmic baryon
density.  Within the last years, a large number of high-quality
observations of hydrogen- and helium- emission lines originating from
extragalactic giant \hii-regions and compact blue galaxies has led to
a number of determinations of the primordial helium mass fraction
$Y_p$ \citep[e.g.][]{peimbert74,kunth83, pagel92, izotov94, izotov97a,
  olive97, izotov98b, peimbert00c}.  Though quoted statistical errors
of these $Y_p$ determinations are often $\lesssim 1$\% due to the size
of the employed samples, it is by now widely accepted that errors in
the inferred $Y_p$ may be dominated by systematic uncertainties. Only
in this way, may one understand how different groups arrive at either
high $Y_p\approx 0.244$ \citep{izotov98b} or low $Y_p\approx 0.234$
\citep{olive97,peimbert00c} values which deviate by several times the
quoted statistical error.

It seems presently feasible to determine the primordial D/H ratio to
unprecedented accuracy by observations of quasar absorption line
systems~\citep{burles99,omeara00,tytler00}. These observations favor a
``low'' primordial deuterium abundance, i.e. D/H $\approx 3\times
10^{-5}$, implying a baryonic contribution to the
critical density of $\Omega_bh^2\approx 0.02$ (with $h$ the Hubble
constant in units of 100 km s$^{-1}$Mpc$^{-1}$), and a primordial
helium abundance of $Y_p\approx 0.247$, within the context of a
standard Big Bang nucleosynthesis scenario.  The accurate
observational determination of $Y_p$ would not only give an
independent \lq\lq measurement\rq\rq of $\Omega_b h^2$, but would also
allow for a test of the validity of a standard Big Bang
nucleosynthesis scenario. Such an undertaking, nevertheless, would
require $Y_p$ determinations with statistical and systematic errors
$\lesssim 1$\% , a magnitude which requires investigation of a large
number of possible errors which may enter the analysis.
 
Beyond the pure random errors that generally underlie any measurement,
there are several sources of systematic uncertainty in commonly used
observational techniques and subsequent analysis of spectra of giant
\hii-regions for the determination of helium (and metal) abundances.
To a first approximation, helium abundances in nebulae can be easily
inferred by observing the relative line fluxes of lines produced
during helium recombinations~\citep{peimbert74} and
lines produced during hydrogen recombinations (e.g. H$\beta$),
\begin{equation}\label{eq:he}
 \frac{I({\rm \hei},\lambda)}{I({\rm H}\beta)}\! =\! \frac{\int\!
  n_{e}n_{\mbox{\tiny \ion{He}{+}}}\alpha_{\mbox{\tiny He}}^{\rm
  eff}(\lambda ,T)}{\int\! n_{e} n_{\mbox{\tiny\ion{H}{+}}}\alpha_{\mbox{\tiny
  H}}^{\rm eff}({\rm H\beta },T)} \propto \frac{\mbox{He}}{\mbox{H}}
\end{equation}
when cloud temperature $T$ and the effective recombination
coefficients $\alpha^{\rm eff}_{i}$ \citep[cf.~][]{osterbrock89} are
known.  The influence of systematic errors starts at the correction
for effects of reddening, underlying stellar absorption (i.e.
correcting for absorption troughs in the stellar continuum at the
position of the emission lines), and fluorescent enhancement of helium
lines (e.g., through absorption of {\hei} $\lambda 3889$ by the
metastable 2$^3$S level of {\hei} and re-emission as {\hei} $\lambda
7065$). The use of Eq.~(\ref{eq:he}) also presumes that there is no
contribution to the observed line radiation from collisional
excitation, an effect which becomes important particularly at higher
densities.  \citet{izotov98b} have devised a method to correct these
potential errors simultaneously by employing several {\hei} emission
lines in the analysis. This method (and others) has also been recently
critically assessed by \citet{olive00c} \citep[see
also~][]{sasselov95}, with the result that residual errors may still
be appreciable.  Uncertainties in the theoretically computed effective
recombination coefficients could also introduce systematic errors as
large as $\sim 1.5\%$ \citep{benjamin99}.

The largest systematic errors, nevertheless, may result from the
idealizing assumptions that \hii-regions are clouds at constant
temperature and density, with simple geometry, and with helium- and
hydrogen- Str\"omgren sphere radii coinciding to within one percent.
These assumptions enter implicitly in an analysis which follows the
strategy of Eq.~(\ref{eq:he}). In this paper the validity of two of
these assumptions are tested in detail: the assumption of constant
temperature (the deviation of this case is commonly referred to as the
existence of \lq\lq temperature fluctuations\rq\rq ) and the
assumption of equal Str\"omgren spheres (correction for this effect is
commonly achieved by factoring in \lq\lq ionization correction factors
(\icf)\rq\rq ).  The latter (\icf) effect is usually only considered
by estimating the quality of the incident continuum by the ``Radiation
Softness Parameter'' $\eta$ \citep{pagel88,skillman89} and based on
model calculations by \citet{stasinska90} it has been assumed that for
sufficiently hard radiation ($\log\eta < 0.9$) \icf-corrections are
negligible ($< 1\%$, \citet{pagel92} ).  Nevertheless, the potential
importance of ionized helium in regions where hydrogen is almost
completely neutral has been recently stressed by a number of groups
\citep{armour99,viegas00,ballantyne00}.  (The problem was already
noted earlier e.g. by \citet{stasinska80}, \citet{dinerstein86}, and
\citet{pena86}.)  New model calculations therein confirmed the
presence of this problem.  The present work is distinguished from
prior analyses by the consideration of a much larger number of model
\hii-regions and incident stellar spectra. It also reaches a somewhat
different conclusion about the importance of this problem.

Effective recombination coefficients $\alpha^{\rm eff}$ employed to
infer ionic abundances from observed emission line ratios are
functions of the electron temperature within the emitting volume.
Temperatures in \hii-regions are usually inferred from flux ratios of
collisionally excited oxygen lines (i.e.
\oiii$\lambda\lambda4959,5007/4363$) and are mostly approximated to be
constant (cf. \citet{peimbert00c} for an analysis going beyond this).
Since collisionally excited lines are exponentially sensitive to
temperature, a temperature determination by such methods is
systematically biased to the hottest parts of the \hii-region.
However, those regions which are somewhat colder may contribute most
of the helium- and hydrogen recombination lines.  A systematically
overestimated electron temperature will also lead to an overestimate
of $Y_{p}$. The problem of the choice of an appropriate mean electron
temperature for recombination coefficients was discussed in detail by
\citet{peimbert67}. A discussion of the temperature effect pertaining
to the determination of $Y_p$ has also appeared in the work of
\citet{steigman97}, who gauge the effect by somewhat arbitrarily
changing inferred temperatures. The present work analyses this problem
by constructing models for \hii-regions with a photoionization code
({\sc Cloudy} 90.05 \citep{hazy}).

The general strategy of this paper is as follows. The simplest
possible, spherically symmetric, \hii-regions at constant (and low)
density are constructed and their emission spectra are calculated
with the help of a photoionization code. These models include a variety of
parameters. Degeneracies of the resultant emission spectra to
these parameters are outlined. Ranges of typical emission line ratios
are inferred from observed \hii-regions, and a model \hii-region is
accepted only if it falls into these ranges. Correction factors $\icf$
and $\tcf$ are defined which quantify the appropriate ionization
correction and correction for temperature inhomogeneity needed to
infer the \lq\lq true\rq\rq\ helium abundance. These factors equal
unity for the ideal ionized, \lq\lq one zone\rq\rq\ cloud.  Finally,
the obtained $\icf$ and $\tcf$ correction factors will be compared to
observational diagnostic tools (i.e. Radiation Softness Parameter and
\oiii\lam5007/\oi\lam6300 ratios) as well as to the \hii-region sample by
\citet{izotov98b}.

In Sect.~\ref{sec:models} the model calculations and the parameters
employed in the calculations are described.  Detailed results 
for the required ionization
correction and temperature correction factors are presented in
Sect.~\ref{sec:uncert} and discussed more broadly in Sect.~\ref{sec:disc}. 
Conclusions are drawn in Sect.~\ref{sec:conclusion}.

\section{Models of \ionsect{H}{ii}-Regions}
\label{sec:models}

The model \hii-regions are calculated using the photoionization code
{\clo} Version 90.05 \citep{hazy}.  In all models a spherical
symmetric distribution of gas around a compact, point-like source of
ionizing radiation at the center is assumed.  All models have a
constant (total) hydrogen number density of $n_{\mathrm{H}}
=10\,$cm$^{-3}$.  This relatively small density is chosen to ensure
that collisional enhancement of helium recombination lines is
negligible. Further, it is assumed that effects of radiative transfer
can be treated in the On-The-Spot approximation.  Thus the emission of
recombination lines may be described using the well-known Case B
limit. Note that the photoionization code Cloudy has been slightly
modified to employ helium emissivities relative to H$\beta$ as given
in \citet{pagel92}. In the limit of small density these are
practically identical to those given in \citet{benjamin99}. The model
clouds investigated here are ionization bounded, i.e. the edge of the
cloud is defined by the Str\"omgren sphere where the degree of
ionization of hydrogen drops rapidly to zero.  In this case all
UV-photons with $h\nu>13.6\,$eV are absorbed and eventually converted
into lower energy photons.

The adopted abundances of heavy elements are those given by
\citet{bresolin99}. These are based on the compilation
by~\citet{grevesse89}, but with depletion on dust grains taken into
account.  In order to simulate \hii-regions with metallicities far
below solar, all elements heavier than helium are scaled down in
abundance by the same factor.  The employed metallicities in the model
calculations range between $Z/Z_{\odot}=1/36$ and $1/12$,
corresponding to $10^6{\rm O/H} = 23.64$ to $70.93$.  The effects of
dust grains, themselves, on, for example, the reddening of lines or the
heating/cooling balance, is not specifically considered.  Assuming
that the density of dust scales with metallicity, these effects should
be fairly small for the clouds considered.  The helium abundance was
kept constant for all models at a value of
$\mbox{He}/\mbox{H}=0.0776$.

\begin{figure}[h]
   \begin{center}
\sidecaption
       \includegraphics[width=.49\columnwidth]{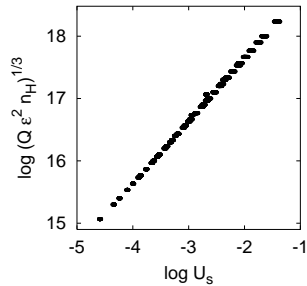}
     \caption{Relationship between $Q$, $\epsilon$, and $U_S$ for 
       a sample of models ionized by a \citet{mihalas72} continuum
       with $\teff=45\,000\,$K. The few models that are more distant
       from the line of proportionality are those where the assumption
       of a solid sphere with $r_{0}\ll r_{S}$ is no longer valid
       (see text).}
     \label{fig:UsQeps}
   \end{center}
\end{figure}

Given the hydrogen number density $n_{\mbox{\tiny H}}$, the spectrum of
the ionizing background, and the metallicity of the cloud, there are
three remaining parameters describing the simulated \hii-regions.
These are $Q(\mbox{H})$ the number of photons with $h\nu>13.6\,$eV
radiated by the source in unit time, $r_{0}$ the inner radius of the
cloud, and $\epsilon$ the volume filling factor of gas at density
$n_{\mbox{\tiny H}}$ taking into account the effect of gas
condensations in \hii-regions \citep[see e.g.][]{osterbrock59}.
Nevertheless, it will be shown that the variation of only one of these
parameters ($Q(\mbox{H})$, $r_0$, or $\epsilon$) is necessary to
generate \hii-regions with different emission spectra.  The quantities
$Q(\mbox{H})$, $n_{\mbox{\tiny H}}$, $\epsilon$, and $r_{0}$ define
the ionization parameter $U_{S}$, i.e. the approximate number of
ionizing photons per hydrogen at the hydrogen Str\"omgren sphere
$r_{S}$:
\begin{equation}
  \label{eq:U_S}
  U_{S}= \frac{Q(\mbox{H})}{4\pi\,r_{S}^{2}n_{\mbox{\tiny H}}c}\, .
\end{equation}
Here the Str\"omgren sphere radius $r_{S}$ may be somewhat arbitrarily
defined as the location where $n_{\mbox{\tiny\ion
    H0}}/n_{\mbox{\tiny\ion H+}}=1$.  The parameter $U_{S}$ fully
determines the structure of a spherically symmetric cloud for a given
chemical composition and incident continuum. Thus clouds with the same
$U_{S}$, but different $ n_{\mbox{\tiny H}}$, $Q(\mbox{H})$, and
$\epsilon$ are self-similar in their ionization structures.  Except
for the total luminosity, such self-similar clouds will result in the
emission of essentially identical nebular spectra.  This assumes that
the effects of collisional excitation on the heating/cooling balance and
the line emission are negligible. This is typically applicable for
densities $n_{\mbox{\tiny H}}\ll 100$cm$^{-3}$.  Moreover, as long as
models have $r_{S}\gg r_0$, the exact choice of the parameter $r_0$ is
of secondary importance for the calculation of nebular emission
spectra.  This is primarily because the regions in the very interior of
the cloud contribute very little to the total line emission due to the
small amount of gas involved.  When $r_{S}\gg r_0$ one can show from
the balance of hydrogen ionizations to recombinations that $U_{S}$ is
proportional to
 \begin{equation}
   \label{eq:U_s-Q-e}
   U_{S}\propto (Q\epsilon^{2}n_{\mbox{\tiny H}})^{1/3}\, .
\end{equation}
Note that this definition of the ionization parameter leads to an
increase of $U_{S}$ with increasing $n_{\mbox{\tiny H}}$ in
contrast to the ionization parameter at the inner radius of the cloud
$U_0=n_{\gamma}/n_{\mbox{\tiny H}}(r_0)$, which is often used as the
input parameter to specify the relative photon to gas density in
plane-parallel clouds.  Figure~\ref{fig:UsQeps} illustrates this
proportionality Eq.~(\ref{eq:U_s-Q-e}) of $U_{S}$ for a sample of models
with varying $Q(\mbox{H})$, $r_0$, and $\epsilon$.  The few models
that do not satisfy Eq.~(\ref{eq:U_s-Q-e}) are those where $r_{S}\sim
r_0$, in particular, where shell geometry pertains.  In what follows,
it is therefore not necessary to vary $Q(\mbox{H})$, $r_0$, and
$\epsilon$ independently, but rather analyze clouds with different
$U_S$, the physical parameter determining the emission spectra when
the geometry of the \hii-regions is characterized by a sphere.

\begin{figure}[h]
  \begin{center}
    \includegraphics[width=\columnwidth]{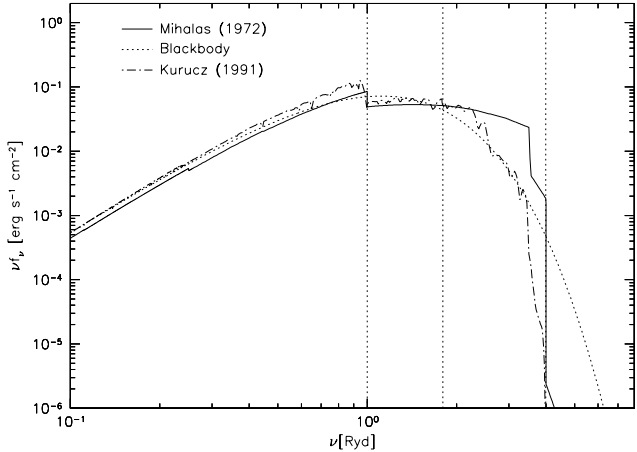}
    \caption{Single star spectra by \protect\citet{mihalas72} and
      \protect\citet{kurucz91}. Both are shown for an effective stellar
      temperature $T_{\mbox{\tiny eff}}=45\,000\,$K and for the same
      number of hydrogen ionizing photons
      $Q(\mbox{H})=10^{51}\,$s$^{-1}$. For comparison, a blackbody
      spectrum with the same parameters is shown as well. Fluxes are
      given at a distance of $73\,$pc to the star. From left to right,
      the vertical lines show the threshold energies for ionization of
      \ion{H}{0}, \ion{He}{0} and \ion{He}{+}.}\label{fig:singlecont}
    \end{center}
\end{figure}

\paragraph{Incident Continuum:}
The effects studied here are particularly sensitive to the shape of
the incident continuum at energies higher than $1\,$Ryd. Spectra with
a larger contribution of helium ionizing photons cause an ionization
structure where the Str\"omgren sphere of helium is equal or even
larger than the Str\"omgren sphere of hydrogen.  These are spectra of O and early B
stars with effective temperatures above $\teff\sim40\,000\,$K.  The
effective temperature of an entire star cluster has these high values
during the early stages of its evolution. In this case the spectrum is
dominated by the most massive and, thus, hottest stars. Since massive stars
burn out more quickly than less massive ones, at later
times, the main contribution to the ionizing spectrum is increasingly
provided by cooler stars. This, in turn, implies a radiation spectrum of the
cluster which is softer. In this paper several different spectra for
the ionizing radiation are used.
These include two different spectra for single stars at various
effective temperatures, in particular, spectra computed from the non-LTE
plane-parallel stellar atmospheres by \citet{mihalas72} and the LTE
plane-parallel atmospheric grids by \citet{kurucz91}.  These continua
are shown in Figure~\ref{fig:singlecont} for one effective
temperature.  For reference, a blackbody spectrum of the same
temperature is also shown in the diagram.  All plotted
spectra are normalized to have the same number of ionizing photons.
From  left to  right, the vertical lines indicate the threshold
energies for ionization of \ion{H}{0}, \ion{He}{0}, and \ion{He}{+},
respectively.
\begin{figure}[h]
  \begin{center}
    \includegraphics[width=\columnwidth]{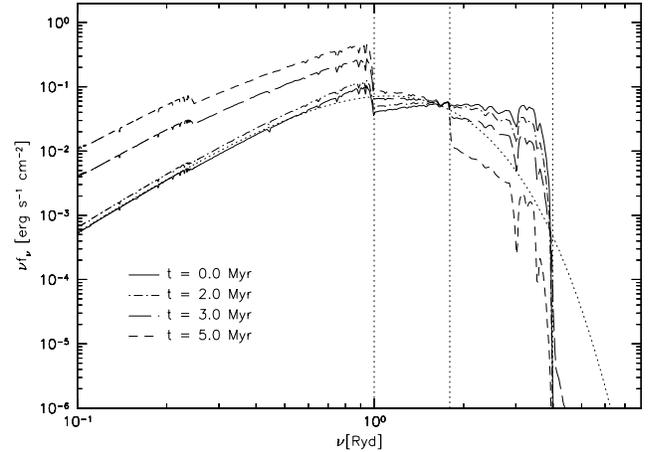}
    \caption{ The continua emitted by stellar clusters of different
      ages with metallicity $Z/Z_{\odot}=1/20$ (see text for details). The
      dotted line shows a blackbody emission with the same effective
      temperature while the vertical dotted lines indicate the
      ionization frequencies of \ion{H}{0}, \ion{He}{0} and
      \ion{He}{+}.}
    \label{fig:sbspec}
  \end{center}
\end{figure}
In addition, this work also employs spectra appropriate to a whole
cluster of stars.  These starburst spectra were generated by using the
starburst synthesis code {\sc Starburst 99} \citep{leitherer99}.  The
synthesis was performed by adopting instantaneous star formation with
Salpeter IMF exponent $2.35$. The population of stars had an upper
mass limit of $150M_{\odot}$ and a fixed stellar mass of
$8.7\times10^{4}M_{\odot}$.  The evolution of the stellar cluster was
described by the standard mass loss tracks with metallicity $Z=0.001$
(i.e.  $Z/Z_{\odot}=1/20$).  The stellar atmospheres needed to
calculate spectra were those from \citet{kurucz92} and
\citet{schmutz92}. A sample of spectra with different ages is shown in
Figure~\ref{fig:sbspec}.  One may see that for longer cluster ages, the
main contribution to the spectrum originates from cooler stars and the
spectrum becomes softer.  Note that in addition to the above mentioned
spectra we have also performed test calculations with Costar stellar
spectra \citet{schaerer97} as well as with spectra given by
\citet{pauldrach98}.  In what follows results for these spectra, which
show generally the same trends as those considered in detail, are not
shown.

\begin{figure}[h]
  \begin{center}
    \includegraphics[width=\columnwidth]{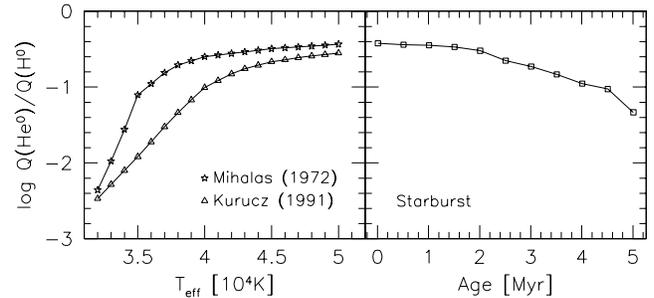}
    \caption{Ratio of the number of helium ionizing
      photons to hydrogen ionizing photons as emitted by different
      stellar continuum sources. The left panel shows the single star
      spectra by \protect\citet{mihalas72} and
      \protect\citet{kurucz91}, whereas the right panel refers to
      continua of a stellar cluster calculated with {\sc Starburst99}
      \protect\citep{leitherer99} for metallicity
      $Z/Z_{\odot}=1/20$}\label{fig:QHeQH}
  \end{center}
\end{figure}

One parameter characterizing the ``hardness'' of the incident
continuum is the ratio of the number of helium ionizing photons
($h\nu> 1.8\,$Ryd) to that of hydrogen ionizing photons
($h\nu>1\,$Ryd), i.e. $Q(\mbox{\ion{He}{0}})/Q(\mbox{\ion{H}{0}})$.
(In addition, the number of \ion{He}{+} ionizing photons can be taken
into account, although in case of the spectra used here, this
contribution is small.)  Figure~\ref{fig:QHeQH} shows this ratio for
the spectra employed in this work. The total luminosity of the
incident continuum is set by specifying the total number of hydrogen
ionizing photons $Q(\mbox{H})$ emitted by the source into $4\pi$.

\paragraph{Constraints from Observations:}
In this work, it is not intended to exactly model one specific cloud,
but rather, to simulate the whole class of low-metallicity
\hii-regions which is employed to infer helium abundances.  Whether or
not a model is suitable to describe the structure and emission of
\hii-regions used for \hel\ determination may be decided by verifying
if the physical parameters adopted for the \hii-regions lead to
numerically calculated spectra which resemble those of the
observations. Therefore, one may choose \lq\lq realistic\rq\rq\ models
according to whether their emission characteristics fall broadly
within certain observationally acceptable ranges.  As a guideline for
typical line emission observed in clouds that are employed for
primordial helium abundance determinations, the sample from
\citet{izotov97a} has been used. Particular importance is placed to
those emission features which are used to infer main properties of the
cloud as well as helium abundances. Ranges for these lines are defined
from the observational sample. They cover the emission characteristics
of the whole sample in these specific lines.  Table~\ref{tab:ranges}
shows these ranges that the emission line ratios of a ``realistic''
model have to be within.

\begin{figure}[h]
  \begin{center}
    \includegraphics[width=\columnwidth]{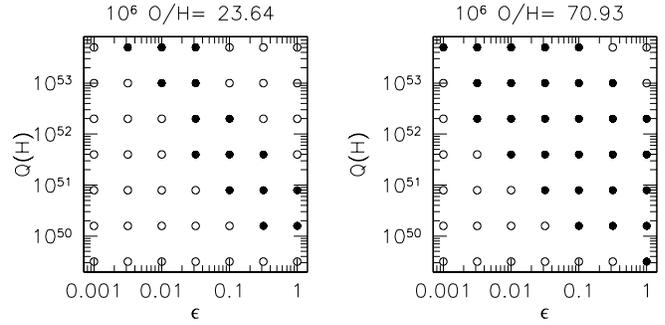}
    \caption{The luminosity $Q(\mbox{H})$ -- filling factor $\epsilon$
      parameter space of \lq\lq acceptable\rq\rq\ models for two different
      metallicities, as labeled. 
      Filled squares indicate the models that fulfill
      the range criterion (see Table~\ref{tab:ranges}) while open
      circles are those where one or more lines are not within the
      observed ranges. All models in this figure were calculated using
      the \protect\citet{mihalas72} spectrum with $T_{\mbox{\tiny
          eff}}=45\,000\,$K.  For other spectra, the structure of the
      diagram stays similar.  The parameter $U_{S}$ stays constant
      along this band structure.}\label{fig:paramplot}
  \end{center}
\end{figure}

In Figure~\ref{fig:paramplot}, each panel shows a grid of models of
different metallicity, where the parameters luminosity $Q({\rm H})$
and filling factor $\epsilon$ are varied, respectively. The filled
dots indicate those models, where the relative emission line
intensities are within the ranges given in Table~\ref{tab:ranges},
while the remaining models are shown by open circles. It is evident
that the acceptable models display a band structure. Models along
these bands are self-similar, with essentially identical emission
spectra but varying total luminosity, in particular, they are
described by the same $U_S$ (cf. Eq.~\ref{eq:U_S}).  In the orthogonal
direction (from the lower left towards the upper right corner) the
ionization parameter $U_{S}$ increases. The same applies for the
temperature within the cloud. Thus, below the band of suitable models,
the line ratios of temperature sensitive emission lines like the
forbidden metal lines, tend to lower values than observed; whereas,
models above the band show emission that is too strong in these lines.
Towards higher metallicity, more models with smaller $U_{S}$ fulfill
the range criterion.  The emission from lower ionization stages of
heavy element ions indicates the significance of ionization fronts at
the outer boundaries of the cloud. These regions become more important
relative to the main body for lower $U_S$, as will be explained below.
\begin{table}[htb]
\begin{center}
  \begin{tabular}[t]{lc|cc}
Ion&\lam$\,$[\AA]&\multicolumn{2}{c}{$I(\lambda)/I(\mathrm{H}\beta)$}\\
  & &min&max\\\hline
\oii\dotfill&  3727 & 0.170 & 4.400\\
Ne$\,$[{\sc iii}]\dotfill& 3869 & 0.130 & 0.890\\
 \hei \dotfill& 3889 & 0.010 & 0.300\\
Ne$\,$[{\sc iii}]+H7\dotfill& 3968 & 0.000 & 0.500\\
H$\delta$\dotfill& 4101 & 0.220 & 0.320\\
H$\gamma$\dotfill& 4340 & 0.410 & 0.620\\
\oiii\dotfill& 4363 & 0.030 & 0.195\\
\hei\dotfill& 4471 & 0.010 & 0.060\\
\heii\dotfill& 4686 & 0.000 & 0.035\\
\oiii\dotfill& 4959 & 0.500 & 2.500\\
\oiii\dotfill& 5007 & 1.500 & 9.200\\
\hei\dotfill& 5876 & 0.007 & 0.130\\
\oi\dotfill& 6300 & 0.000 & 0.500\\
\siii\dotfill& 6312 & 0.000 & 0.037\\
H$\alpha$\dotfill& 6563 & 2.540 & 3.100\\
\hei\dotfill& 6678 & 0.007 & 0.050\\
\sii\dotfill& 6717 & 0.019 & 0.500\\
\sii\dotfill& 6731 & 0.014 & 0.374\\
\hei\dotfill& 7065 & 0.010 & 0.040\\\hline
\end{tabular}
  \caption{Range criteria for the line emission in wavelength
    $\lambda$ to be fulfilled by the model
    \hii-regions. Fluxes $I$ are given relative to H$\beta$. 
    The intervals refer to the
    values that are observed by \protect\citet{izotov97a}. Some limits are
    relaxed by 3-10\% to take into account possible deviations of real
    clouds from the idealized assumptions of the model calculations.}
  \label{tab:ranges}
\end{center}
\end{table}

\section{Results of the Model Calculations}
\label{sec:uncert}

\subsection{Ionization Structure}
The determination of helium abundances in \hii-regions possibly
requires a correction for unseen ionization stages of hydrogen or
helium.  When the stars ionizing the nebulae are not too hot (late O
stars and later), the supply of helium-ionizing photons might not be
sufficient to maintain a high fraction of \ion{He}{+} throughout the
whole hydrogen Str\"omgren sphere. Vice versa, for hard spectra of the
ionizing source it is also possible that a helium Str\"omgren sphere
larger than the one of hydrogen is set up
\citep{stasinska80,dinerstein86,pena86,armour99,viegas00,ballantyne00}.
Recently, it has been attempted to correct for this effect, i.e. the
existence of neutral hydrogen \citep{viegas00,ballantyne00}, though
these studies reach opposite conclusions about the induced systematic
uncertainty on the inferred $Y_p$.
 
Another possible systematic error due to the ionization structure of
\hii-regions arises from the fact that the observed emission lines are
always weighted towards regions with higher electron density. The
intensity in a particular \hei{} recombination line relative to a
reference line like H$\beta$ is given by
\begin{equation}
  \frac{I(\mbox{\hei},\lambda)}{I(\mbox{H}\beta)} =
  \frac{\int\!n_{e}n_{\mbox{\tiny\ion{He}{+}}} \alpha_{\mbox{\tiny
  He}}(\lambda ,T)dV}{
  \int\!n_{e}n_{\mbox{\tiny\ion{H}{+}}}\alpha_{\mbox{\tiny
  H}}(\mbox{H}\beta ,T)dV}   
\label{eq:Emisabu}
\end{equation}
Here $n_{e}$ is the electron number density, $n_{\mbox{\tiny
    \ion{He}{+}}}$ and $n_{\mbox{\tiny \ion{H}{+}}}$ are the number
densities of \ion{He}{+} and \ion{H}{+}, respectively, and
$\alpha_i(\lambda ,T)$ are the recombination coefficients of the lines
$\lambda$ considered as a function of electron temperature $T$.  Thus,
in regions where the electron density is small, the emission of a line
becomes weaker even if the density of the emitting ion itself stays
constant.  If $n_{e}$ is not constant within the whole volume, though,
the intensity ratio in Eq.~(\ref{eq:Emisabu}) is not proportional to
$\ion{He}{+}/\ion{H}{+}$ even when the electron temperature and, thus
recombination coefficients $\alpha$, stay constant over the cloud
volume.  This effect becomes particularly important for clouds where
the Str\"omgren sphere of hydrogen is smaller than the one for helium.
This is due to the small abundances of all other elements, in
particular, \ion{H}{+} provides the main contribution to $n_{e}$.  But
also in case of a \ion{He}{+} sphere smaller than the \ion{H}{+}
sphere this effect needs to be considered. The coupling of the
ionization equilibrium equations of helium and hydrogen for photons
with $h\nu\geq 24.6\,$eV, causes variations in $n_{e}$ resulting in
smaller {\hei} recombination line emission \citep{sasselov95}.

In order to correct the observationally inferred helium abundances for
unseen ionization stages, usually an ``ionization correction factor''
\icf{} is introduced. The significance of this effect, is then
estimated by comparing certain emission lines that provide information
on the incident continuum and, in turn, on the ionization structure.
In contrast to the definition of \icf{} that is used by
  \citet{viegas00}, \citet{armour99} and others, \icfx{} as it is
  introduced here, corrects for both, unseen helium ionization stages
\ion{He}{0} and \ion{He}{2+}, and the effect arising from varying
electron density within the observed volume
\citep[see][]{stasinska90}. The definition of the ionization
  correction factor by \citet{stasinska80}, that is employed in the
  work of \citet{pagel92} and \citet[ and subsequent
  papers]{izotov94}, takes the electron density into account. However,
  in these works \icfx{} is assumed to be unity in most cases. (See
  the discussion later.)  In the present work \icfx{} is defined as
the ratio of the true helium abundance relative to hydrogen and the
amount of ionic helium to hydrogen within the volume $V$ that is
emitting line radiation at all:
\begin{eqnarray}
  \label{eq:icftotal}
  \icfx &=& \left(\frac{\int n_{\mbox{\tiny He}}dV}{\int n_{\mbox{\tiny
  H}}dV}\left/ \frac{\int n_{\mbox{\tiny\ion{He}{+}}}dV}{\int n_{\mbox{\tiny
  \ion{H}{+}}}dV}\right.\right)\times\nonumber\\&\times&\left( \frac{\int n_{\mbox{\tiny\ion{He}{+}}}dV}{\int n_{\mbox{\tiny
  \ion{H}{+}}}dV}\left/\frac{\int n_{e}n_{\mbox{\tiny \ion{He}{+}}}dV}{\int n_{e}n_{\mbox{\tiny\ion{H}{+}}}dV}\right.\right) \nonumber\\[2ex]
    &=& \frac{\int n_{\mbox{\tiny He}}dV}{\int n_{\mbox{\tiny
  H}}dV}\left/\frac{\int n_{e}n_{\mbox{\tiny \ion{He}{+}}}dV}{\int n_{e}n_{\mbox{\tiny\ion{H}{+}}}dV}\right.
\end{eqnarray}
Note that the inclusion of correction for \ion{He}{2+} in
Eq.~(\ref{eq:icftotal}) may be omitted, as the abundance of this ion
may be observationally inferred from $\lambda 4686$ line radiation. In
the models used here the abundance of doubly ionized helium is
essentially negligible, such that both definitions would coincide.
The first factor in the first equation of Eq.~(\ref{eq:icftotal}) may
be identified as the ionization correction for constant electron
  density. It is important to realize that the effect of varying
electron density, taken into account by the second factor in this
equation, implies a required correction which, in almost all cases, is
significantly smaller (i.e. $\left|(\icfx-1)/(\icf - 1)\right| < 1$)
than the one where the effect is omitted.

\subsubsection{General Trends:}
\label{sec:icf}

\begin{figure}[ht]
   \begin{center}
     \includegraphics[angle=-90,width=\columnwidth]{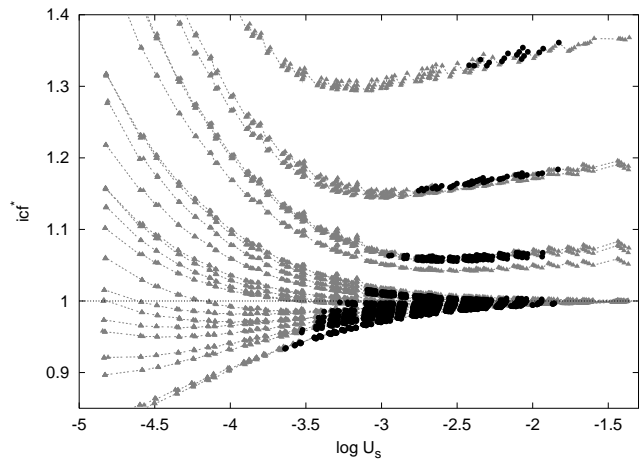}
     \caption{Ionization correction factors as a function of 
       ionization parameter at the
       Str\"omgren sphere $U_{S}$. Dark symbols refer to
       models fulfilling the range criteria. 
       All models employed in this work are shown.}
     \label{fig:Us-icfnew}
   \end{center}
\end{figure}

In the following, general trends of the required ionization correction
factors for clouds with varying parameter $U_{S}$ and for ionizing
sources of different spectra are studied. Figure \ref{fig:Us-icfnew}
shows the relation between $U_{S}$ and \icfx{} for the whole sample of
models. The models that fulfill the range criterion discussed above,
are shown as black filled symbols.  Each track in this figure
corresponds to a grid of models ionized by the same continuum. The
lower tracks result from relatively hard spectra (i.e. high \teff)
while for softer spectra \icfx{} tends towards larger values.
(Continua were chosen in steps of 1000\,K or 0.5\,Myr, respectively.)
Furthermore, it may be seen that when the spectrum of the ionizing
source is fixed \icf-corrections generally become more pronounced for
smaller ionization parameter $U_{S}$.
It is evident from the figure that systematic errors induced by the
ionization structure may, in principle, grow quite large. Even for
models, that show emission line characteristics similar to those
observed in \hii-regions employed for $^{4}$He determinations (black
symbols) this potential error ranges from an overestimate of $^{4}$He
by up to $\sim 7$\% to an underestimate by up to $\gtrsim30$\%. It
will be shown below, however, that existing observational data on
low-metallicity \hii-regions seems consistent with a possible $^{4}$He
overestimate, rather than a $^{4}$He underestimate.

\paragraph{Influence of the Incident Continuum:}
\begin{figure}[htb]
  \begin{center}
    \includegraphics[angle=-90,width=\columnwidth]{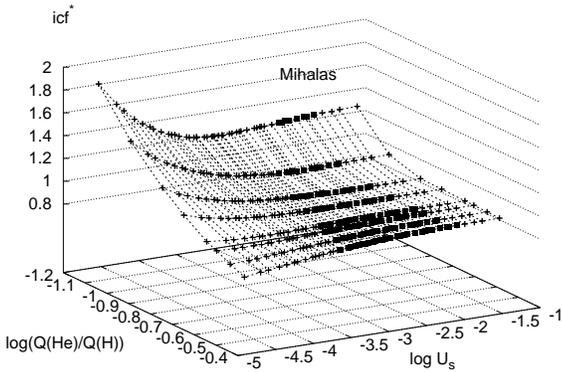}
    \caption{\icfx{} versus $Q(\mbox{\ion{He}{0}})/Q(\mbox{\ion H0})$
      and $U_{S}$ for clouds ionized by \protect\citet{mihalas72} spectra.
      Dark squares refer to models fulfilling the emission line
      range criterion.}\label{fig:QQUsicfmih}
  \end{center}
\end{figure}
To study the influence of the incident continuum on the ionization
corrections in more detail, subsamples were analyzed.
Fig.~\ref{fig:QQUsicfmih} shows the derived \icfx{} of models ionized
by different spectra as a function of the
$Q(\mbox{\ion{He}{0}})/Q(\mbox{\ion{H}{0}})$-ratio and the ionization
parameter $U_{S}$. Though the particular figure employs Mihalas
spectra for the ionizing radiation, the general trends of the figure
are hardly changed when other ionizing spectra are employed. This is
mainly due to the choice of more physical
$Q(\mbox{\ion{He}{0}})/Q(\mbox{\ion{H}{0}})$-ratios as characteristics
of the spectra, rather than the effective stellar temperatures.  It
may be seen that for clouds ionized by relatively soft spectra (small
values of $Q(\mbox{\ion{He}{0}})/Q(\mbox{\ion{H}{0}})$) the \icfx\ is
larger than unity for all $U_{S}$. When the hardness of the spectrum
is increased, the \icfx\ correction approaches unity until a critical
value for $Q(\mbox{\ion{He}{0}})/Q(\mbox{\ion{H}{0}})$ is reached. For
values larger than about
$Q(\mbox{\ion{He}{0}})/Q(\mbox{\ion{H}{0}})\gtrsim 0.15$ required
ionization corrections reverse such that values of \icfx\ smaller than
unity obtain. This trend is observed to be more dramatic for smaller
$U_{S}$, where the \icfx\ may deviate significantly from unity. From
an observational point of view it would be desirable to infer helium
abundances within clouds with $U_{S} \gtrsim 10^{-2}$ and
$Q(\mbox{\ion{He}{0}})/Q(\mbox{\ion{H}{0}})\gtrsim 0.2$ where at least
within the simple models analyzed here icf-corrections are within a
per cent.

Qualitatively, these trends may be understood as follows: For soft
spectra the supply of He ionizing UV-photons emitted from the central
source is not sufficient to ionize He throughout the \ion H+ sphere
regardless of the ionization parameter. This leads to $\icfx>1$.  Only
photons with $h\nu\geq 24.6\,$eV may ionize helium, whereas all
photons with $h\nu\geq 13.6\,$eV may ionize hydrogen. In the ``best''
case scenario (for maximum helium ionization of a radiation bounded
nebula) all $h\nu\geq 24.6\,$eV will be absorbed by helium such that
the ionization equilibria of hydrogen and helium may be treated
separately. Then for \icfx{}$>1$ the relation
\begin{equation}
\icfx \propto\frac{n_{\mbox{\tiny He}}}{n_{\mbox{\tiny
    H}}}\left(\frac{Q(\mbox{\ion{He}{0}})}{Q(\mbox{\ion{H}{0}})}\right)^{-1}
\end{equation}
approximately holds. This expression neglects the emission of hydrogen
ionizing radiation during helium recombination, as well as assuming
$Q(\mbox{\ion{He}{0}})\ll Q(\mbox{\ion{H}{0}})$.  It, nevertheless,
clarifies the existence of a critical
$Q(\mbox{\ion{He}{0}})/Q(\mbox{\ion{H}{0}})$ where \icfx{} tends
towards corrections larger than unity, irrespective of $U_{S}$.  This
critical value depends on the helium content relative to hydrogen.

There exists an additional effect which leads to $\icf^*$ deviating
from unity, even when there is a sufficient supply of helium ionizing
photons.  For decreasing $U_{S}$, the width of the Str\"omgren sphere
$\delta r_{S}$ increases relative to the Str\"omgren radius $r_{S}$.
The width $\delta r_{S}$ is given by the distance over which ionizing
radiation in partially neutral gas has optical depth of order unity,
i.e.  $\tau \approx \sigma_{ph}\,\epsilon\, n_{\rm H}\,\delta
r_{S}\approx 1$.  Here $\sigma_{ph}$ is an appropriate photoionization
cross section.  From this one may show that $\delta r_{S}/r_{S}$
relates to $U_{S}$ via
\begin{equation}   
\delta r_{S}/r_{S} \approx 5\times 10^{-7}\,U_{S}^{-1}
\end{equation}
as long as $r_{S}\gg r_{0}$, and under the assumption of constant
density across the ionization front.  Since the relative volume of gas
in the transition region to that in the essentially completely ionized
body of the cloud is $\sim 3\delta r_{S}/r_{S}$ the effects of
emission from partially ionized gas become increasingly important for
small $U_{S}$.  This implies significant ionization corrections for
small $U_{S}$, overestimating helium abundances for
$Q(\mbox{\ion{He}{0}})/Q(\mbox{\ion{H}{0}})\gtrsim 0.15$ and
underestimating helium abundances in the opposite case.  Thus \icf\ 
effects become asymptotically unimportant only for spectra that have
sufficient emission of helium ionizing photons and for Str\"omgren
sphere ionization parameters which are sufficiently large.

\subsubsection{Observational Tools}
Because neither the ionization parameter $U_{S}$ itself nor the ratios
of different $Q$-values are directly observable, other quantities have
to be found in order to estimate the influence of ionization structure
on the inferred abundances. 

\begin{figure}[ht]
   \begin{center}
     \includegraphics[angle=-90,width=\columnwidth]{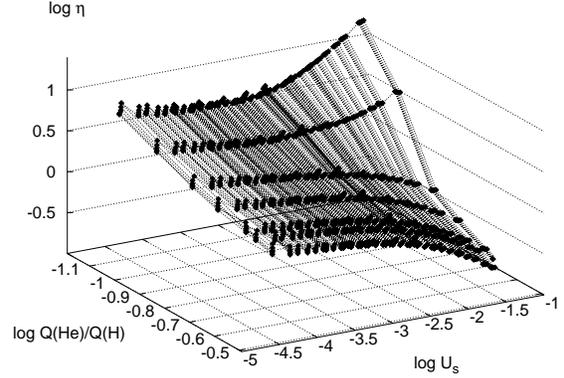}
     \caption{Dependency of the Radiation Softness Parameter $\eta$ on the
       ionizing photon $Q(\mbox{\ion{He}{0}})/Q(\mbox{\ion H0})$ ratio
       and the ionization parameter $U_{S}$. This figure shows a
       sample of models ionized by \protect\citet{mihalas72} continua
       of different effective temperatures. }
     \label{fig:Us.eta.mih}
   \end{center}
\end{figure}

\paragraph{Radiation Softness Parameter:}
One frequently chosen method compares the abundance ratios \ion
O+/\ion{O}{++} and \ion{S}{+}/\ion{S}{++}.  This provides a measure
for the quality of the incident spectrum with respect to its ionizing
ability.  This ``Radiation Softness Parameter''
\begin{equation}
  \label{eq:etatr}
  \eta  =\frac{\ion{O}{+}}{\ion{S}{+}}
  \frac{\ion{S}{++}}{\ion{O}{++}},  
\end{equation}
originally introduced by \citet{pagel88}, is often used to ``read
off'' the ionization correction from model calculations like those
performed by \citet{stasinska90}.  Usually, first the particular ionic
abundances are inferred from their respective emission lines, then
$\eta$ is derived from these results following the definition of
Eq.~(\ref{eq:etatr}). Here the particular choice of emission lines
employed for the ionic abundance determinations may differ slightly
between different authors \citep[see e.g.~][]{pagel92}.  In many
observational determinations of helium abundances, the ionization
correction for He is simply assumed to be negligible for \hii-regions
that are ionized by sufficiently hard spectra leading to small values
of $\eta$. For instance, \citet{pagel92}  adopt no correction  
  (i.e. $\icfx{}\equiv1$) for those nebulae where $\log\eta<0.9$ and
exclude objects with larger $\eta$ from their analysis.
\citet{izotov98b}, more or less, follow this procedure.  Corrections
$<1$ are usually not considered. The $\log\eta$ of the objects which
this group uses for helium determinations, are all within the interval
of $-0.2<\log\eta<0.4$.  Figure~\ref{fig:Us.eta.mih} shows the values
for $\eta $ as a function of the ionization parameter $U_{S}$ and the
ratio $Q(\mbox{\ion{He}{0}})/Q(\mbox{\ion{H}{0}})$.  From this figure
it is evident that $\log\eta$ has not only a dependence on the quality
of the spectrum, as often assumed, but also on the parameter $U_{S}$.
\hii-regions which would be ideally suitable for helium abundance
determinations would have extremely low $\log\eta < -0.5$, since,
following the discussion from above, they would have large $U_{S}$ and
$Q(\mbox{\ion{He}{0}})/Q(\mbox{\ion{H}{0}})$ above the critical value.
Unfortunately, such regions seem not to be easily found in
observational surveys. Note here that clouds illuminated by
\citet{kurucz91} spectra typically show larger $\log\,\eta$ for the
same $Q(\mbox{\ion{He}{0}})/Q(\mbox{\ion{H}{0}})$ and $U_{S}$ than
those illuminated by either \citet{mihalas72} or {\sc Starburst 99}
spectra.  This shift is mainly caused by the generally much smaller
emission of photons that ionize \ion{O}{+} ($h\nu=2.58\,$Ryd) in
\citet{kurucz91} spectra.
\begin{figure}[h]
  \begin{center}
    \includegraphics[angle=-90,width=\columnwidth,clip=true]{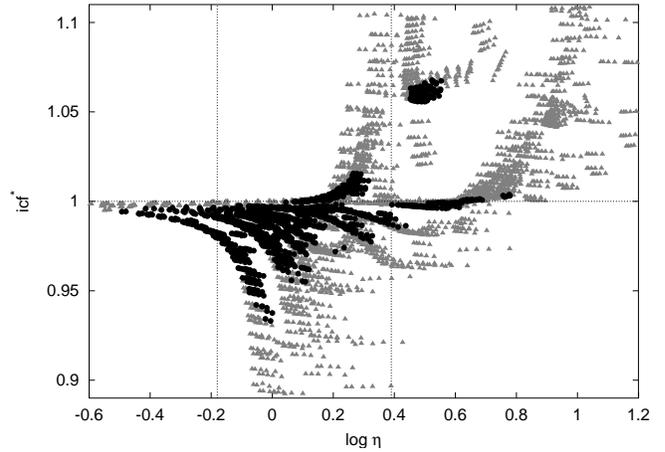}
    \caption{Derived values of \icfx{} versus the Radiation Softness
      Parameter $\eta$. The vertical lines indicate the range of
      $\eta$ values  of the sample investigated in  \citet{izotov97a}. }
    \label{fig:eta-icf}
  \end{center}
\end{figure}

Figure \ref{fig:eta-icf} shows the calculated \icfx{} versus the
Radiation Softness Parameter $\eta$. The $\eta$ of the \hii-regions
from the sample of \citet{izotov97a} are within the interval that is
indicated in this figure by the two vertical lines. the figure shows,
that $\icfx{}$ may become significantly smaller than unity leading to
a potential overestimate of the helium abundance in these clouds of up
to $\sim6$\% if no correction is applied.

\begin{figure}[htb]
  \begin{center}
  \includegraphics[angle=-90,width=\columnwidth]{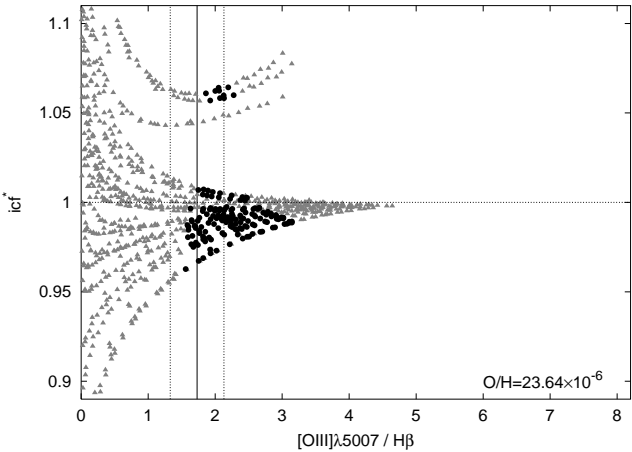}
  \caption{Metallicity dependent cutoff criterion suggested by
    \protect\citet{ballantyne00}. Shown are the derived values for
    {\icfx} versus the emission of \oiii\lam5007 relative to H$\beta$
    for models with low metallicity. The vertical lines indicate the
    suggested cutoff and its error. }
  \label{fig:OHb-icf.zlo} 
\end{center} 
\end{figure}

\begin{figure}[htb]
  \begin{center}
  \includegraphics[angle=-90,width=\columnwidth]{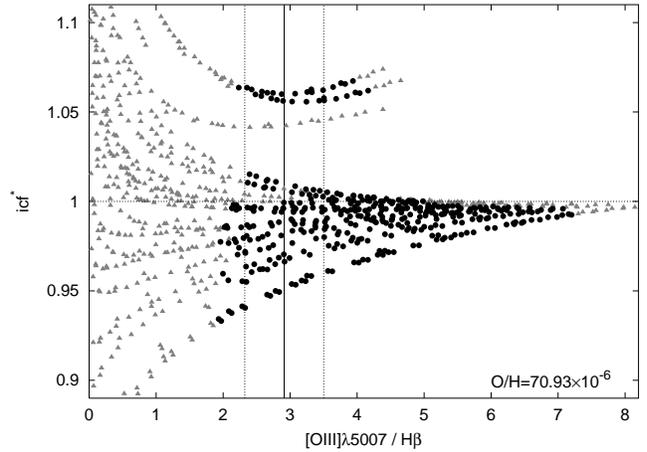}
  \caption{Same as Fig.~\ref{fig:OHb-icf.zlo} but for models with high
    metallicity.}
  \label{fig:OHb-icf.zhi} 
\end{center} 
\end{figure}

\begin{figure}[htb]
  \begin{center}
  \includegraphics[angle=-90,width=\columnwidth]{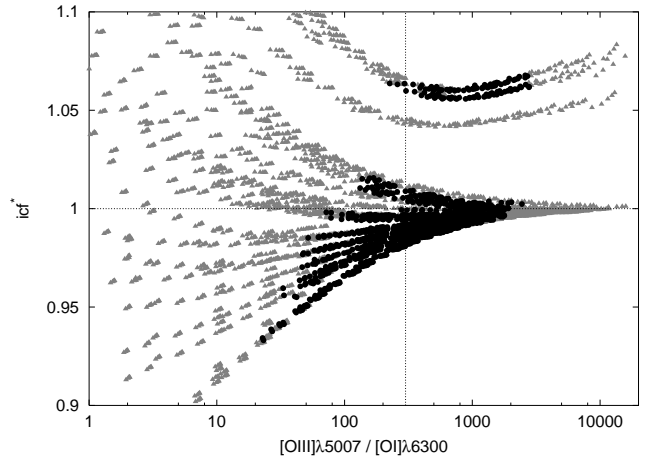}
  \caption{Metallicity independent cutoff criterion by
    \protect\citet{ballantyne00}. This shows the \icfx{} versus the
    line ratio \oiii\lam5007/\oi\lam6300. The cutoff value is marked
    by the dashed line.}
  \label{fig:Orat-icf} 
\end{center} 
\end{figure}

\paragraph{Oxygen-line cutoff-criterion:}
More recently \citet{ballantyne00} suggested two interesting criteria
possibly suitable for an estimate of the significance of ionization
corrections. The first one involves the \oiii\lam 5007 line relative
to H$\beta$ and, thus, depends on the metallicity. The second
criterion employs the ratio \oiii\lam5007/\oi\lam6300 and should be
approximately independent of metallicity:
\begin{equation}
\left(\frac{\mbox{\oiii\lam}5007}{\mbox{\oi\lam}6300}\right)_{\!\!\mbox{\tiny
      cutoff}}\!\!= 300
\end{equation}and 
\begin{eqnarray}
  \label{eq:cutoff}
\lefteqn{\left(\frac{\mbox{\oiii}\lambda5007}{\mbox{H}\beta}\right)_{\!\!\mbox{\tiny
      cutoff}}\!\!=\nonumber}\\
&=&(0.025\pm0.004) \left(\frac{\mbox{O}}{\mbox{H}}\right)
      \times 10^{6}+(1.39\pm0.306)
\end{eqnarray}
They state that all objects observed with emission ratios lower than
these cutoff values should be excluded from consideration since they
may be subject to large ``reverse'' ionization corrections.  Figures
\ref{fig:OHb-icf.zlo} to \ref{fig:Orat-icf} show the results of our
calculations in terms of these criteria. It may be seen that the
potential error due to $\icf<1$ at the cutoff is still significant and
may reach values up to 5\%.  From these results it is obvious, that
the suggested cutoffs are still too optimistic and even larger values
have to be employed. Unfortunately, there is a scarcity of observed
\hii-regions which have sufficiently large $\lambda 5007/\lambda 6300$
or $\lambda 5007/{\rm H\beta }$ to reduce \icf{}-corrections to a
small magnitude.  The main difference between our models and the
models calculated by \citet{ballantyne00} is the geometry. The \icf{}
corrections tend towards larger values in spherical geometry as the
outer regions of the cloud where preferably lower degrees of
ionization are found, contribute more to the total emission in a
particular line than in the plane parallel case.  Nevertheless, our
disagreement with the paper of~\citet{ballantyne00} on the required
cutoff values for negligible \icf{}-correction is not due to geometry.
Rather, the proposed cutoffs by \citet{ballantyne00} are not
sufficient to exclude potentially large reverse icf effects
\citep{ferland00pr}.

\subsection{Temperature Variations}
\label{sec:tempvar}

In a second focus of this paper, uncertainties are investigated that
may occur from the commonly used technique to obtain a suitable
electron temperature $T_e$ for helium and hydrogen recombination
coefficients.  Even in homogeneous, constant density model nebulae,
the temperature shows variations (i.e. a temperature stratification),
whereas the abundance estimates employing emission lines usually
assume a constant temperature. Following prior literature we sometimes
refer to this temperature stratification as the existence of
temperature \lq fluctuations\rq.  The variations in $T_{e}$ arise because
the thermal equilibrium depends strongly on the local abundance of
different ionization stages of coolants as, for example, oxygen.  In
real nebulae temperature variations could be even more pronounced due
to, for example, more complicated density structures, possibly
inducing \lq real\rq\ small-scale fluctuations in the temperature.
Variations of $T_{e}$ are typically only taken into account for the
abundance estimate of metals, i.e.  oxygen.  This is accomplished by
dividing the observed \hii-region into two zones of different degree
of ionization adopting different temperatures inferred from
appropriate ions for each zone.  The main problem in dealing with
nebular temperatures is to find the appropriate mean temperature for
the physical process considered.  Since the line emissivities for
different lines and ionization stages of different elements have
varying temperature dependencies, the bulk of the emission of certain
lines used for analysis may originate from either within hotter or
cooler regions than the helium line emission.  This implies that, in
order to obtain accurate results for each line, a different
temperature, determined by a appropriate weighted average, would need
to be employed \citep{peimbert67}.  This, however, is a far from 
straightforward task (cf. \citet{peimbert00c}).  Usually, the electron
temperature influencing the emissivities of helium lines is assumed to
be constant within the \ion{He}{+} sphere.  It is determined by
observing the ratios between the flux of collisionally excited oxygen
lines \oiii\lam\lam5007,4959 and \lam4363 \citep{aller84}.
Nevertheless, it is known that such lines are exponentially
temperature sensitive and thus mostly originate within the hotter
regions of the nebulae.

In order to quantify the potential uncertainty arising from this
method of temperature determination, another correction factor \tcf{}
is introduced by
\begin{eqnarray}
  \label{eq:deftcf}
  \lefteqn{\tcf =\frac{\displaystyle\frac{\alpha_{\mbox{\tiny He}}
(\lambda ,T_{\mbox{\tiny\oiii}})}{\alpha_{\mbox{\tiny H}}
({\rm H}\beta ,T_{\mbox{\tiny\oiii}})}
  \frac{\int n_{e}n_{\mbox{\tiny\ion{He}{+}}}dV}
{\int n_{e}n_{\mbox{\tiny\ion H+}}dV}}{\displaystyle%
\frac{\int n_{e}n_{\mbox{\tiny\ion{He}{+}}}\alpha_{\mbox{\tiny He}}
(\lambda ,T)dV}
{\int n_{e}n_{\mbox{\tiny\ion H+}}\alpha_{\mbox{\tiny H}}
({\rm H}\beta ,T)dV}}}\\
  &&=\frac{\int n_{e}n_{\mbox{\tiny
  \ion{He}{+}}}dV}{\int n_{e}n_{\mbox{\tiny
  \ion{H}{+}}}dV}\left/\left(\frac{N_{\mbox{\tiny
  \ion{He}{+}}}}{N_{\mbox{\tiny \ion{H}{+}}}}\right)_{\!\!T_{\mbox{\tiny
          \oiii}}}\right.\nonumber
\end{eqnarray}
where
\begin{equation}
\label{eq:extra}
\left(\frac{N_{\mbox{\tiny
  \ion{He}{+}}}}{N_{\mbox{\tiny \ion{H}{+}}}}\right)_{\!\!T_{\mbox{\tiny
          \oiii}}} =
\frac{\alpha_{\mbox{\tiny H}}(T_{\mbox{\tiny\oiii}})}{\alpha_{\mbox{\tiny
  He}}(T_{\mbox{\tiny\oiii}})}
  \frac{I(\mbox{He\,I},\lambda)}{I(\mbox{H}\beta)} \nonumber
\end{equation}
Here $T_{\mbox{\tiny\oiii}}$ denotes the ``average'' temperature as
derived from the oxygen lines. The \tcf{} factor is unity for constant
temperature within the whole emitting volume. For ``normal'' clouds
this factor is smaller unity since the emission of \oiii{} lines
originates in regions with high amount of \ion{O}{2+} that do not
coincide with the region where helium is ionized. Thus, this emission
is weighted towards the hotter interior of the cloud. In contrast the
main contribution to helium recombination lines is emitted within the
outer, and thus cooler, parts of the cloud.  Note, however, that in
extreme cases the \tcf{} may become larger than unity which is due to
it's implicit dependence on ionization structure (cf.
Eq.~\ref{eq:deftcf}). In such cases, nevertheless, the clouds are
already characterized by extreme \icf{} corrections.

\begin{figure}[htb]
  \begin{center}
   \includegraphics[angle=-90,width=\columnwidth]{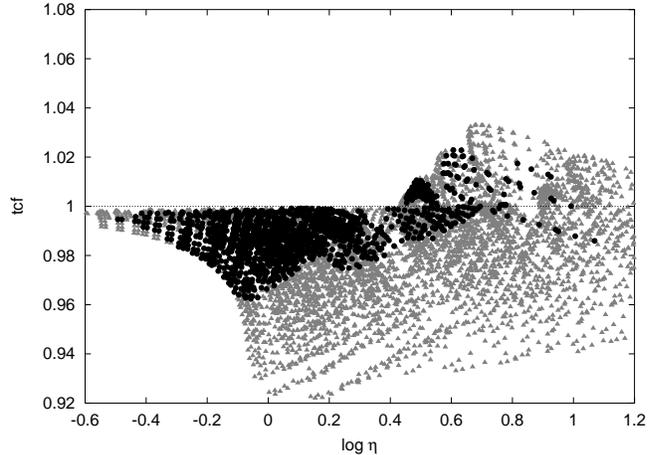}
    \caption{Temperature correction factor \tcf{} 
    versus the Radiation Softness Parameter $\eta$ for the whole
    sample of models computed.}
    \label{fig:eta-tcf-all} 
\end{center} 
\end{figure}

Corrections due to temperature variations depend on the particular
helium line considered since emissivities of different lines have
different temperature dependencies.  Observers sometimes employ the
\hei\,\lam6678 line alone because it is the line that is less suspect
of enhancement due to collisional excitation than other \hei\ lines.
Other approaches use a weighted mean of all strong \hei\ 
recombination lines. Of the commonly used \hei\ recombination lines
\lam 4471, \lam5876, and \lam6678, the last should be most sensitive
to potentially arising effects from temperature variations.  Figure
\ref{fig:eta-tcf-all} shows the \tcf{} versus $\eta$ for all models
computed in this work. It is derived as an equally weighted average of
the \tcf{}'s of the individual \lam 4471, \lam5876, and \lam6678
lines.  Because the metal lines provide the main cooling contribution,
varying temperature effects are not only a function of ionization
structure but also depend strongly on metallicity. Thus, the tracks in
Figure \ref{fig:eta-tcf-all} are broader than those found for the
\icfx{} corrections. From these calculation one may estimate a
potential systematic overestimate of $Y$ due to a non-ideal choice of
temperature of up to 4\%.

In case of strongly varying electron density, the effects of
ionization structure and temperature variation are not anymore
independent because higher order terms introduce a cross correlation
between both effects. Since \icf{}-effects, in some cases, may be
considerably larger than \tcf{}-effects this cross correlation may
lead to unexpectedly high values for the ``ideal'' temperature
yielding $\tcf{}>1$. Such models are characterized by extreme
ionization structures.  They certainly do not fulfill the simplified
assumptions of distinct regions of high and low degree of ionization.
Interestingly, however, even those regions show emission properties
that may be observed in real clouds.

\begin{figure}[htb]
  \begin{center}
    \includegraphics[angle=-90,width=\columnwidth]{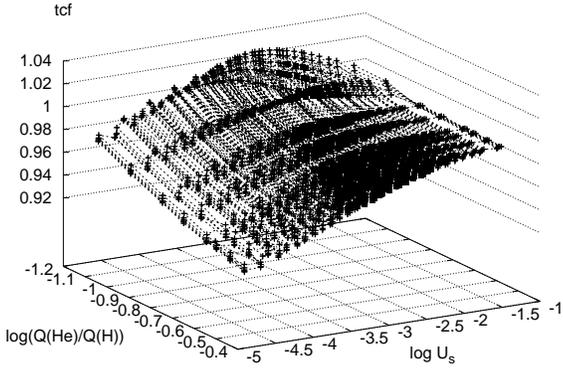}
    \caption{\tcf\, corrections versus the
      $Q(\mbox{\ion{He}{0}})/Q(\mbox{\ion{H}{0}})$ ratio and $U_{S}$
      for models ionized
      by Mihalas continua.}
    \label{fig:QQUstcf}
  \end{center}
\end{figure}
In Fig.\ref{fig:QQUstcf}\, temperature correction factors are shown as
a function of $U_{S}$ and the quality of the incident continuum as
given by the ratio $Q(\mbox{\ion{He}{0}})/Q(\mbox{\ion{H}{0}})$. The
shown models have been computed with Mihalas spectra though results
are very similar when different spectra are employed. It is evident
that the \tcf{} has it's strongest dependence on $U_{S}$ and is only
mildly dependent on the spectrum.  Those models which have \tcf{}'s
close to unity have large $U_{S}$ and
$Q(\mbox{\ion{He}{0}})/Q(\mbox{\ion{H}{0}}) \gtrsim 0.15$ above the
threshold for sufficient helium ionization. This is very similar to
the behavior of \icf{}-effects as discussed in the last section, such
that both potential errors could be reduced if \hii-regions of only
very small radiation softness parameter $\log\eta < -0.5$ would be
used for helium abundance determinations.

\section{Discussion}\label{sec:disc}
\begin{figure}[htb]
  \begin{center}
    \includegraphics[angle=-90,width=\columnwidth]{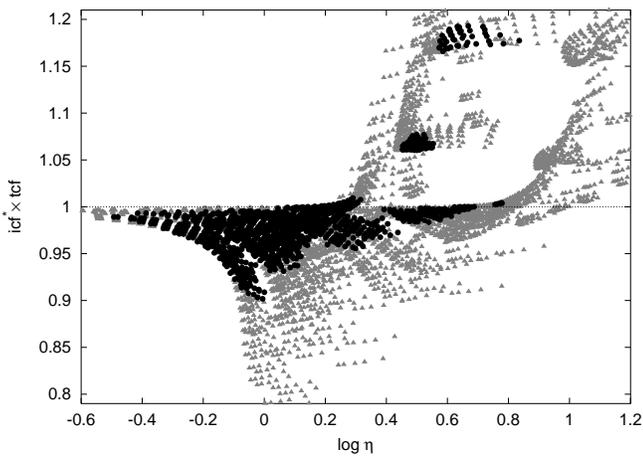}
    \caption{The total required correction factor (\icfx\ and {\tcf})
      versus the Radiation Softness Parameter $\eta$ for all models 
      considered in this work.}\label{fig:eta.tcficf}
  \end{center}
\end{figure}

From the results of the preceding section it is clear that potentially
large uncertainties in the determination of helium abundances in
low-metallicity \hii-regions may arise due to the ``imperfect''
ionization structure of nebulae, as well as the existence of
temperature variations. Moreover, both of the investigated systematic
uncertainties point towards a potential overestimate of helium
abundances, since observed \hii-regions seem to be described by
radiation softness parameters $\eta$ which are indicative of possible
``reverse'' ionization corrections, rather than showing evidence for
additional amounts of neutral helium. Fig.~\ref{fig:eta.tcficf} shows
the combined required correction factor, taking account of both
effects, as a function of $\eta$. As before, models which fulfill the
emission line range criteria are indicated by black symbols.

Corrections for ionization structure seem to depend mostly on two
parameters, the spectrum of the incident radiation, as well as the
ratio of width to radius $\delta r_{S}/r_{S}$ of the Str\"omgren
sphere. As long as the supply of helium ionizing photons is sufficient
(i.e. $Q(\mbox{\ion{He}{0}})/Q(\mbox{\ion{H}{0}})\gtrsim 0.15$
corresponding to $\log\,\eta \lesssim 0.3$ within the assumed models)
only ``reverse'' ionization corrections ($\icf{} < 1$) occur. These
ionization corrections could be, at least in principle, reduced to
negligible levels when the ionization parameter at the Str\"omgren
sphere becomes large, corresponding to small $\delta r_{S}/r_{S}$.
This limit occurs for $\log\,\eta \lesssim -0.5$ for the model
\hii-regions employed here.  Such corrections could also become small
when matter-bounded nebulae are considered.  In the intermediate
regime $-0.5 \lesssim \log\,\eta \lesssim 0.3$ \icf{} corrections may
become appreciable with the required correction being larger for
harder spectra.  Non-uniform temperature corrections, parameterized by
\tcf{}, depend mostly on $\delta r_{S}/r_{S}$ with the dependence on
spectrum of secondary importance. These corrections also have some
dependence on metallicity, as it is known that low-metallicity clouds
have shallower temperature gradients. Both corrections individually
become most significant for radiation softness parameters $\log\,\eta
\approx 0$ leading to the apparent ``triangular'' structure in
Fig.~\ref{fig:eta.tcficf}.

Ionization correction factors in helium abundance determinations have
been recently investigated by \citet{armour99, viegas00} and
\citet{ballantyne00}. The present work broadly confirms the potential
problem of required ``reverse'' ionization corrections, though their
are differences on the viability of proposed criteria to exclude such
uncertainties as well as the magnitude of the uncertainties when
variations in electron density is taken into account. Moreover,
\citet{ballantyne00} infer an underestimate of the primordial helium
abundance due to this effect, whereas \citet{viegas00} argue for only
a moderate $\sim 1\%$ overestimate.

\citet{steigman97} concluded that the existence of temperature
``fluctuations'' in \hii-regions leads to an underestimate of the
primordial helium abundance. This underestimate is claimed to be the
result of an underestimate of metal abundances, which leads to a
decreased slope of helium abundance against metallicity, as well as to
an overestimate of collisional enhancement of helium recombination
lines. Whereas the present analysis has not studied the first effect,
their is disagreement on the sign of the second effect.
\citet{steigman97} lower the temperatures of \hii-regions from the
inferred $T_{\mbox{\tiny\oiii}}$ by a typical amount of $\Delta T
\approx 2000\,$K, in order to account for a systematic overestimate of
temperature from collisionally excited oxygen lines.  An overestimate
of $\sim 20\%$ in temperature leads to an overestimate of $\sim 5\%$
in the helium abundances since the temperature dependence of the ratio
of recombination emissivities $\alpha_{\mbox{\tiny
    H}}/\alpha_{\mbox{\tiny He}}$ (cf. Eq.~\ref{eq:extra}) is $\sim
T^{0.25}$ for $\lambda 6678$, $\sim T^{0.23}$ for $\lambda 5876$, and
$\sim T^{0.13}$ for $\lambda 4471$, respectively.  Coincidently, a
value of $\sim 5\%$ is close to the largest \tcf{} corrections found
in this work. This correction may be compared to the potential error
in derived collisional enhancement factors (taken from
\citet{kingdon-ferland95}) due to a possible overestimate of
temperature. Most clouds in the sample of \citet{izotov98b} have $n_e
< 100\,$cm$^{-3}$. Assuming $n_e = 100\,$cm$^{-3}$ this change is
$\sim 0.3\%$ for $T_{\mbox{\tiny\oiii}}\approx 10^4\,$K and $\sim 1\%$
for $T_{\mbox{\tiny\oiii}}\approx 2\times 10^4\,$K. Only for clouds of
fairly large density $n_e \approx 300\,$cm$^{-3}$ and temperature
$T_{\mbox{\tiny\oiii}}\approx 2\times 10^4\,$K does the ``collisional
error'' become comparable $\sim 2.5\%$ to the error induced by
employing the wrong temperature in the recombination coefficients. It
is concluded that an overestimate of temperature leads in most cases
to an overestimate of helium abundance.

\begin{figure}[htb]
  \begin{center}
    \includegraphics[width=\columnwidth]{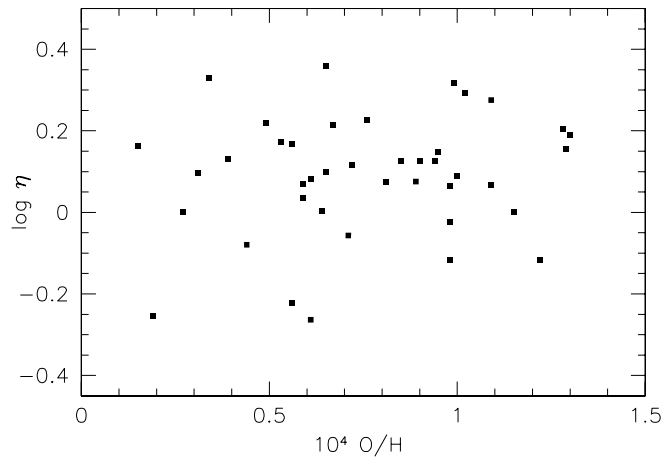}
    \caption{Distribution of Radiation Softness Parameter $\eta$ versus
      metallicity in terms of the oxygen to hydrogen ratio for the
      sample by \protect\citet{izotov98b}.}
    \label{fig:etaOH}
  \end{center}
\end{figure}

\begin{figure}[htb]
  \begin{center}
    \includegraphics[width=\columnwidth]{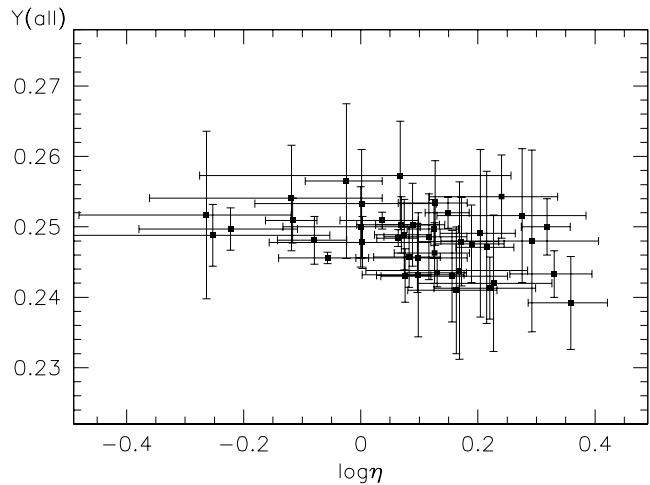}
    \caption{Inferred helium mass fraction versus the Radiation
      Softness Parameter $\eta$ for the sample of compact blue
      galaxies observed by
      \protect\citet{izotov98b}.}\label{fig:conclude1}
  \end{center}
\end{figure}

Though the employed model \hii-regions in this paper are probably
still too simplistic to account for the structure of real
\hii-regions, it would be interesting to establish or refute whether
existing observational data is consistent with systematic
uncertainties of sign and magnitude as found in the previous section.
In Fig.~\ref{fig:conclude1} helium abundances of the sample by
\citet{izotov98b} are shown as a function of the radiation softness
parameter as inferred by these authors.  It is striking to observe
that \hii-regions with $\log\,\eta \gtrsim 0$ do populate the area of
lower helium $Y\sim 0.24$ abundance, whereas their lower $\eta$
counterparts seem not to. The approximate transition region of this
trend at $\log\,\eta \approx 0$ coincides with the peak of the
triangular structure observed in Fig.~\ref{fig:eta.tcficf}, where the
potentially largest correction factors were found. Note that the
helium abundances shown in Fig.~\ref{fig:eta.tcficf} are not corrected
for stellar enrichment.  One may wonder if the apparent correlation in
this graph is caused by a correlation of the inferred $\eta$ of
\hii-regions with cloud metallicity. In Fig.~\ref{fig:etaOH} radiation
softness parameter is shown as a function of metallicity, illustrating
that such a correlation does not exist. An explanation of the observed
trend in Fig.~\ref{fig:conclude1} due to the existence of composite
clouds, with one cloud illuminated by a hard spectrum ($\icf{}\approx
1$) and another by a soft spectrum ($\icf{} > 1$), seems also
unlikely, since composite clouds should rarely have $\log\,\eta
\lesssim 0.4$~\citep{viegas00}.

\begin{figure}[htb]
  \begin{center}
    \includegraphics[angle=-90,width=\columnwidth]{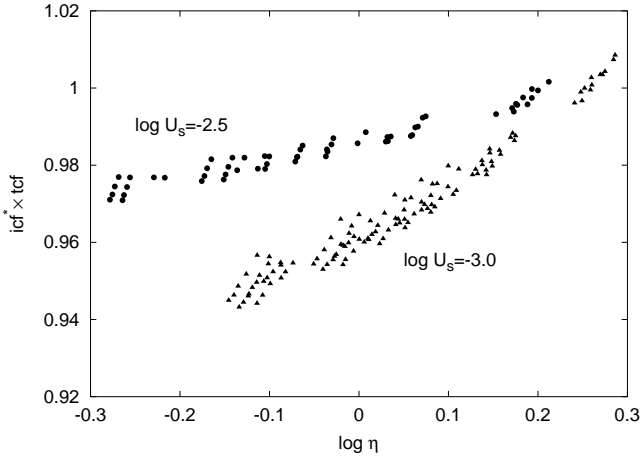}
    \caption{Resulting total correction $\tcf\times\icfx$ as a function
      of $\log\eta$ for models with (approximately) fixed $U_{S}$, as
      labeled.  Shown are models illuminated by starburst and Mihalas
      spectra. All models fulfill the range criteria.}
    \label{fig:conclude3}
  \end{center}
\end{figure}

Though it is not clear how statistically significant the excess of low
$Y$ \hii-regions for $\log\,\eta \gtrsim 0$ is, these findings seem
suggestive that the investigated systematic uncertainties may indeed
exist in the observational data.  One may verify upon inspection of
Figures~\ref{fig:Us-icfnew}, \ref{fig:Us.eta.mih}, and
\ref{fig:QQUstcf} that a sample of clouds illuminated by the same hard
spectra, but with varying $U_{S}$ (i.e. $\delta r_{S}/r_{S}$), may not
explain such a trend.  Nevertheless, the same figures (and the
knowledge that lower branches in Figure~\ref{fig:Us-icfnew} correspond
to larger $Q(\mbox{\ion{He}{0}})/Q(\mbox{\ion{H}{0}})$, i.e. harder
spectra) illustrate that a sample of clouds with typical $\log\,U_{S}
\sim -3$ illuminated by spectra of varying hardness may easily explain
the observed trend. This is illustrated in Fig.~\ref{fig:conclude3}
where the combined correction factors are shown for a sample of models
illuminated by \citet{mihalas72} and starburst spectra, and for two
particular values of $U_{S}$. If this indeed would be the explanation,
helium abundances could be overestimated by a typical $\sim 4\%$, with
required corrections in individual cases possibly as large as $\sim
8\%$.

\section{Conclusions}
\label{sec:conclusion}

The primordial helium abundance is commonly inferred from
observationally determined helium abundances within low-metallicity
extragalactic \hii-regions. Such observational determinations may be
subject to systematic uncertainties due to required corrections for
existence of neutral hydrogen or helium (\icf{}-corrections) and due
to the possibility of non-uniform cloud temperature (referred to as
\tcf{}-corrections).  These effects have been investigated in this
paper.  A chain of correction factors has been constructed that leads
from the directly observable fluxes of helium- and hydrogen-
recombination lines, and the inferred temperature from collisionally
excited oxygen lines, to the true He/H ratio (cf. Eq.~\ref{eq:extra})
in a nebula:
\begin{equation}\label{eq:alleffects}
  \frac{\displaystyle N_{\mbox{\tiny He}}}{\displaystyle N_{\mbox{\tiny
  H}}}= \icfx\!\times\!\tcf\!\times\!\left(\frac{N_{\mbox{\tiny
  \ion{He}{+}}}}{N_{\mbox{\tiny \ion{H}{+}}}}\right)_{\!\!T_{\mbox{\tiny
          \oiii}}}
\end{equation}
A large number of spherically symmetric model \hii-regions with
varying parameters and illuminated by ionizing radiation of different
spectra have been constructed with the help of a photoionization code.
It was shown that such models may be generally characterized by three
physical quantities: the number of helium- to hydrogen- ionizing
photons $Q(\mbox{\ion{He}{0}})/Q(\mbox{\ion{H}{0}})$, metallicity, and
the ionization parameter at the Str\"omgren sphere $U_{S}$. It was
also shown that $U_{S}^{-1}$ is proportional to the ratio of width to
radius $\delta r_{S}/r_{S}$ of the Str\"omgren sphere. The computed
models are compared to the emission characteristics of observed
\hii-regions.  A critical value for
$Q(\mbox{\ion{He}{0}})/Q(\mbox{\ion{H}{0}}) \gtrsim 0.15$ has been
identified, such that for spectra above this critical value only
reverse \icf{} corrections apply. Such spectra result in clouds with
radiation softness parameter $\log\,\eta \lesssim 0.3$.  Reverse
\icf{} corrections may lead to a potentially large (up to $\sim 6\%$)
overestimate of helium abundances, particularly for \hii-regions of
small $\log\, U_{S}$. These uncertainties may not easily be excluded
even when \oiii\lam5007/\oi\lam6300 -ratios \citep{ballantyne00} are
considered. In addition, a possible overestimate of helium abundances
of similar magnitude (up to $\sim 4\%$) may be present due to the
existence of temperature gradients, provided that temperature is
inferred from collisionally excited oxygen lines.  Both effects are
absent only for \hii-regions of extraordinarily small radiation
softness parameter $\log\,\eta \lesssim -0.5$, corresponding to clouds
of small $\delta r_{S}/r_{S}$ (large $U_{S}$) illuminated by
sufficiently hard spectra. In contrast, the range of $-0.3 \lesssim
\log\,\eta \lesssim 0.3$ is found for an existing
sample of \hii-regions~\citep{izotov98b}. Here required corrections
may become large.  This sample seems to display a correlation between
inferred helium abundance and radiation softness parameter in
concordance with a typical overestimate of helium by about $\sim 2 -
4\%$.  In case such an interpretation of the data should prevail, and
when uncertainties leading to a possible underestimate of helium
abundances due to other effects would be excluded, a downward revision
of inferred primordial helium would be required.

\begin{acknowledgements}
The authors wish to acknowledge T.~Abel and S.~Burles for several
helpful discussions and G.~Ferland for communication.
\end{acknowledgements}

\end{document}